\documentclass[runningheads]{llncs}
\usepackage{graphicx}
\usepackage{cite}
\usepackage[table,xcdraw]{xcolor}
\usepackage{tabularx}
\usepackage{caption}
\usepackage{subcaption}
\usepackage{tabularx}
\usepackage{adjustbox}
\usepackage{multirow}

\newcolumntype{L}[1]{>{\raggedright\let\newline\\\arraybackslash\hspace{0pt}}m{#1}}
\newcolumntype{C}[1]{>{\centering\let\newline\\\arraybackslash\hspace{0pt}}m{#1}}
\newcolumntype{R}[1]{>{\raggedleft\let\newline\\\arraybackslash\hspace{0pt}}m{#1}}
\begin{document}
%
\title{Individual Behavioral Insights in Schizophrenia: A Network Analysis and Mobile Sensing Approach}
\titlerunning{Individual Behavioral Insights in Schizophrenia}
%
\author{Andy Davies\inst{1}\orcidID{0009-0009-5807-1972} \and
Eiko Fried\inst{2}\orcidID{0000-0001-7469-594X} \and
Hane Aung\inst{1}\orcidID{0000-0002-4376-0849} \and
Omar Costilla-Reyes\inst{3}\orcidID{0000-0001-8331-7262}}

\authorrunning{A. Davies et al.}
%

\institute{School of Computer Sciences, University of East Anglia, Norwich, NR9 7TJ, UK
\email{\{andy.davies, Min.aung\}@uea.ac.uk}\\
\and
Faculty of Social Sciences, Institute of Psychology, Leiden University, Rapenburg 70, 2311, Netherlands
\email{e.i.fried@fsw.leidenuniv.nl}
\and
Computer-Aided Programming Research Group, MIT Computer Science and Artificial Intelligence Laboratory (CSAIL), MA, 02139, USA
\email{costilla@mit.edu}}

\maketitle              
\vspace{-5mm}
\begin{abstract} 
Digital phenotyping in mental health often consists of collecting behavioral and experience-based information through sensory and self-reported data from devices such as smartphones. Such rich and comprehensive data could be used to develop insights into the relationships between daily behavior and a range of mental health conditions. However, current analytical approaches have shown limited application due to these datasets being both high dimensional and multimodal in nature. This study demonstrates the first use of a principled method which consolidates the complexities of subjective self-reported data (Ecological Momentary Assessments - EMAs) with concurrent sensor-based data. In this study the CrossCheck dataset is used to analyse data from 50 participants diagnosed with schizophrenia. Network Analysis is applied to EMAs at an individual (n-of-1) level while sensor data is used to identify periods of various behavioral context. Networks generated during periods of certain behavioral contexts, such as variations in the daily number of locations visited, were found to significantly differ from baseline networks and networks generated from randomly sampled periods of time. The framework presented here lays a foundation to reveal behavioural contexts and the concurrent impact of self-reporting at an n-of-1 level. These insights are valuable in the management of serious mental illnesses such as schizophrenia.

\end{abstract}

\section{Introduction}
Schizophrenia is a complex, Serious Mental health Illness (SMI) that develops in approximately 1\% of the global population \cite{saha2005systematic} and represents a significant personal and economic burden at an individual, familial and societal level \cite{chong2016global, wander2020schizophrenia}. Symptoms can include hallucinations (both visual and auditory), disordered and delusional thinking, impaired cognitive ability, disorganized speech and behavior \cite{patel2014schizophrenia}, as well as increased social isolation, withdrawal and amotivation \cite{mccutcheon2020schizophrenia}. Although characterised as a chronic condition, the disease course is not static, with diagnosed individuals typically fluctuating between periods of partial remission and periods of symptomatic relapse \cite{wang2017predicting, national2014psychosis, emsley2013nature}. Studies have identified symptomatic and behavioral changes that can manifest prior to relapse \cite{ascher2010cost, emsley2013nature, birchwood2000schizophrenia, he2020assessing}, however, these changes often remain undetected until the occurrence of significant negative consequences \cite{wang2016crosscheck}. Evidence further suggests that timely clinical intervention poses an effective strategy in the prevention of further deterioration, and the transition into a state of full relapse \cite{morriss2013training, wang2016crosscheck}. 

\begin{figure}
    \centering
    \includegraphics[width=\textwidth]{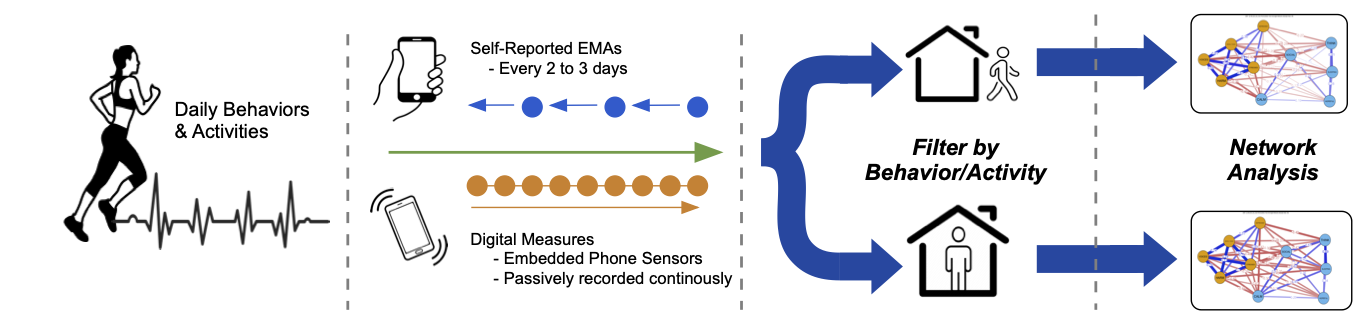}
    \caption{Overview of the processes involved in this study; networks are generated using EMA responses concurrent with specific sensor-based behavioral contexts.} \label{fig1}
\end{figure}

In this paper, we seek to demonstrate a method which consolidates the complexities of subjective self-reported Ecological Momentary Assessments (EMAs) when accounting for variations in behavioral context. Using the CrossCheck dataset, a first of its kind dataset combining real-world, longitudinal behavioral data, and self-reported EMAs specific to schizophrenia \cite{wang2016crosscheck}; we aim to demonstrate the effectiveness of using sensor-based data to identify periods of various behavioral context, from which network analysis can be applied to observe and compare differences in network connectivity and the relationships between corresponding EMAs. Specifically, we focus on behaviors that can be categorised according to periods of sociability and social isolation, both noted symptoms associated with schizophrenia symptom severity \cite{he2020assessing}.  Figure \ref{fig1} provides a high-level overview of this process, from individual-level (n-of-1) data through to behavioral filtering and network analysis. Ultimately, the goal of this framework is to reveal behavioral contexts and their resulting impact on self-reported EMAs at an n-of-1 level, in particular providing insights into symptomatic improvement or disease exacerbation.

\vspace{-5mm}
\vspace{-10mm}

\section{Related Work}\label{sec:relate}
Conventional research into human behavior often relies on time and resource intensive data collection through face-to-face engagement in a controlled or clinical environment. However, the pervasiveness of mobile technology \cite{silver2019smartphone} in everyday life is affording researchers and clinicians access to vast quantities of moment-by-moment, in-situ, individual-level data captured by personal digital devices; the granular level quantification of which is referred to as digital phenotyping \cite{torous2016new, mohr2017personal}. These personal phenotypes \cite{barnett2018relapse, onnela2021opportunities} provide a digital fingerprint from which psychological, cognitive and behavioural characteristics can be measured and assessed \cite{davidson2022crossroads, insel2018digital, insel2017digital, mohr2020digital}; providing valuable insights into symptomatic markers and effective psychiatric treatments \cite{perez2021wearables}. Within mental health research, digital phenotyping has been employed in a number of studies, including student mental health \cite{melcher2020digital}, depression \cite{kamath2022digital}, prediction of suicidal urges \cite{brown2022digital}, anxiety disorders \cite{jacobson2022digital}, social anxiety \cite{jacobson2020digital}, and psychosis spectrum illnesses \cite{benoit2020systematic}. 

With an increasing emphasis on patient-centred healthcare and individualised medicine \cite{bradbury2020practice}, digital phenotyping lends itself to move away from the population level \cite{fisher2018lack} and instead conduct n-of-1 trials (or single subject trials); these focus on an individual patient as the sole unit of observation throughout a study \cite{lillie2011n}. Typically, n-of-1 studies have been used within both clinical and research settings to assess pharmaceutical efficacy and treatment viability within individual participants \cite{punja2016n}. The focus of these trials enables the identification of observations or characteristics that may not be evident in a collective population-level analysis. However for larger population samples, the insights gained from n-of-1 trials can contribute to larger-scale Randomized Control Trials (RCTs).

The CrossCheck collection emulated a Randomized Control Trial (RCT) design \cite{chalmers1981method} that explored the viability of continuous remote patient monitoring through a multimodal sensing system; the core aim of which sought to accurately predict indicators of symptomatic and psychotic relapse in Schizophrenia Spectrum Disorders (SSD) \cite{wang2016crosscheck}. CrossCheck's digital phenotyping dataset identified unique digital indicators of psychotic relapse; for some participants changes in self-reported EMAs provided actionable descriptors of symptom exacerbation, whilst in others, passively recorded behavioural and sensory data proved useful in identifying changes in established behaviors or daily functioning \cite{ben2017crosscheck}. A recent study demonstrated the detection of decreases in symptoms using change-point algorithms and counterfactual explanations \cite{canas2023counterfactual}. Additional research using the CrossCheck dataset mapped features on a two-dimensional space using t-Distributed Stochastic Neighbor Embedding (t-SNE), a technique used for dimensionality reduction that projects each high-dimensional data point to a two-dimensional data point \cite{van2008visualizing}. Using t-SNE, CrossCheck visualized data points that represented a participant's behavioural features used to predict EMA responses; when plotted, these data points clustered according to each specific study participant. This clearly demonstrated that there are observable differences between study participants and that CrossCheck's sensor data is highly person dependent \cite{wang2016crosscheck}. At a population level, the initial study found significant associations between recorded behavioral features and changes in mental health indicators; in particular decreased levels of physical activity and sociability was associated with negative mental health, whilst improvements in established sleep patterns and getting up earlier collated with positive mental health \cite{wang2016crosscheck}. These findings were further supported through research into behavioral stability using the same dataset, this stability index drew on participant's passively recorded features and behaviours to assess the extent to which a diagnosed participant adheres to a stable routine. This study identified correlations between the stability index of recorded features and symptomatic severity, the results of which demonstrated that greater periods of stability in social activities - such as calls and SMS messages - was associated with reduced symptoms. In contrast, increased stability in periods of inactivity - time spent still - exhibited an association with increased symptom severity \cite{he2020assessing}. The findings of these studies highlight not only the highly person-centric nature of CrossCheck's multimodal data, but also the close association between daily behaviors and symptom severity; both of which are of particular importance to this study as we seek to analyse the impact recorded behavioral contexts have on self-reported EMAs at an individual level.

In recent years the use of network analysis within psychological research has become an important tool in the estimation and visualization of psychological data, and can be used to identify multivariate patterns and relationships \cite{borsboom2021network, hevey2018network}. Within these networks, nodes represent variables such as mood states collected via EMAs, with edges between nodes denoting statistical relationships between said nodes \cite{hevey2018network}. The process by which these associations and relationships are calculated can vary depending on the initial dataset and selected statistical model \cite{borsboom2021network}. In recent years, network analysis techniques have been applied to a large number of datasets in order to gain deeper insights into a range of mental health problems, including drug and alcohol dependency \cite{rhemtulla2016network}, suicidal behavior in adolescents \cite{fonseca2022risk}, depression and anxiety \cite{beard2016network}, and the treatment of psychosis \cite{bak2016n}. Network analysis consists of three stages; network structure estimation, network description, and network stability analysis \cite{borsboom2021network}. Network structure estimation refers to the process by which the underlying structure of a network is inferred, involving the selection of relevant nodes and edges as well as selecting an optimal statistical model. Network description is the characterisation of a network which involves understanding network topology and node centrality. Finally, network stability analysis refers to the examination of a network's robustness, consistency and the accuracy of edge weights \cite{borsboom2021network}. Recently, estimating network models on time-series data with numerous repeated observations using EMA data has gained traction, however this presents three distinct challenges. First, most work is estimated at the group level, ignoring potential variation across participants in network structures. Second, networks are stationary, i.e., one network is obtained throughout the time period, assuming network structure does not vary by context. Third, it is unclear how network analysis ought to deal with multimodal (e.g., sensor and EMA) data. Here we tackle all three challenges, by estimating n-of-1 networks according to a selected behavioral (sensor-based) context prior to EMA network generation, as a result enabling more nuanced, context-dependent qualitative networks.

\vspace{-2mm}
\section{Dataset}\label{sec:data}
The CrossCheck \cite{wang2016crosscheck} dataset originally consisted of $150$ participants, each of whom met the criteria for schizophrenia as defined in the DSM-IV \cite{american1994diagnostic} and DSM-V \cite{american2013diagnostic}, whilst also meeting CrossCheck's inclusion criteria \cite{ben2017crosscheck, wang2016crosscheck}. Organised into two groups of $75$, participants within the CrossCheck study arm were each issued with a smartphone that continuously recorded a range of behavioral, sensory and self-reported EMA data over a 12 month period \cite{wang2016crosscheck, ben2017crosscheck, wang2020social}. Embedded smartphone sensors passively record daily behaviours and activities continuously, whilst EMAs were self-reported every 2 to 3 days \cite{wang2016crosscheck}. Of particular relevance to this paper are the following features:

\vspace{-1mm}
\paragraph{Ecological Momentary Assessment (EMA):}
EMAs afford a viable way of capturing real-time psychological data within a natural environment. Every 2 to 3 days, CrossCheck administered a 10-item self-reporting assessment designed to measure schizophrenia-related thoughts, feelings, and behaviors \cite{ben2017crosscheck, wang2016crosscheck}. Each question (see Table \ref{tab:emas}) was answered on a scale from 0 ("Not at all") to 3 ("Extremely"). For ease of analysis and understanding, each EMA is grouped according to either its positive or negative association.

\vspace{-7mm}
\begin{table}[]
    \centering
    \caption{CrossCheck EMA Questions}
    \label{tab:emas}
    \resizebox{0.9\textwidth}{!}{%
        \begin{tabular}{ L{6cm} | L{6cm} }
            \textbf{Positive EMAs} & \textbf{Negative EMAs} \\ \hline
            \rowcolor[HTML]{D9D9D9} 
            + Have you been feeling \textbf{CALM}? 
                & - Have you been \textbf{DEPRESSED}? \\
            + Have you been \textbf{SOCIAL}? 
                & - Have you been feeling \textbf{STRESSED}? \\
            \rowcolor[HTML]{D9D9D9} 
            + Have you been \textbf{SLEEPING} well? 
                & - Have you been bothered by \textbf{VOICES}? \\
            + Have you been able to \textbf{THINK} clearly? 
                & - Have you been \textbf{SEEING THINGS} other people can’t see? \\
            \rowcolor[HTML]{D9D9D9} 
            + Have you been \textbf{HOPEFUL} about the future? 
                & - Have you been worried about people trying to \textbf{HARM} you?
        \end{tabular}%
    }
\end{table}
\vspace{-7mm}

\paragraph{Behavioural Sensing:}
Study smartphones continuously collected a wide range of behavioral features for each participant, however only the following behavioral features are relevant to this paper due to their close association with sociability and social isolation. - \textit{Geo-spatial Activity:} refers to timestamped locational data derived using a combination of device GPS, Wifi and cellular network towers \cite{wang2016crosscheck, ben2017crosscheck}. - \textit{Speech Frequency \& Duration:} periods of human speech was inferred from ambient sound using the inbuilt device microphone \cite{ben2017crosscheck}. - \textit{Calls \& SMS:} The frequency and duration of incoming and outgoing calls, as well as the number of incoming and outgoing SMS messages is passively logged and recorded by the CrossCheck application \cite{wang2016crosscheck, ben2017crosscheck}.

\vspace{-1mm}
\section{Method}
The following section presents the proposed methodology employed in this study, from prerequisite data pre-processing, through to initial network generation and statistical analyses. Data pre-processing requires basic resampling and thresholding, with network generation simply based on the correlations within the 10 set EMA questions (see Table \ref{tab:emas}). As such giving low computational requirements and ease of re-implementation.

\vspace{-1mm}
\subsection{Data Pre-processing}
From CrossCheck's original study arm, $50$ participants have been identified from which further analysis can be conducted, these individuals were selected for the quantity and quality of their recorded data. Participants whose engagement was limited, or who recorded inconsistent and unusable data were omitted from further analysis. Whilst both sensory and device usage data is temporally continuous, self-reported EMA responses are only given every 2 to 3 days \cite{wang2016crosscheck}. However, each EMA question also pertains to days prior to a given response, as such we can retrospectively replicate each EMA score to also be concurrent with sensor data recorded between EMA responses. Any days that fall outside of the 2 day back fill window are omitted from a participant's dataset.

Selection of sensor features was based on their ability to effectively capture defined behavioral contexts, without the need for further processing to map sensor data to a particular context (e.g. it is a reasonable assumption that no location data outside of the primary residence indicates not leaving the home, and that no calls or detected conversations indicates not verbally socializing).  Table \ref{tab:features} lists these selected features and their corresponding categories. These categories are created based on 2 factors; first relating to the veracity of the behavioral context that the data represents as stated above, with the second factor being based on well understood behavioral contexts for this SMI, such as social isolation and sociability. Along side these features random sampling of unfiltered data is also conducted from which an empirical distribution is generated, this serves as a baseline from which comparisons can then be made.

\vspace{-7mm}
\begin{table}[!ht]
\caption{Selected behavioral contexts and their corresponding categories.}
\label{tab:features}
\centering
\resizebox{\textwidth}{!}{%
\begin{tabular}{L{5cm}|C{4cm}|C{4cm}}
\textbf{Behavioral Feature in a 24hr period} 
    & \textbf{Periods of Social Isolation} 
        & \textbf{Periods of Sociability} \\ 
\hline
\rowcolor[HTML]{E0E0E0} 
Baseline
    & \textit{Random Unfiltered Sample} 
        & \textit{Random Unfiltered Sample} \\ 
Locations Visited
    & \textit{No locations visited} 
        & \textit{$\geq 1$ locations visited} \\ 
\rowcolor[HTML]{E0E0E0} 
Calls Made
    & \textit{No calls made} 
        & \textit{$\geq 1$ calls made} \\ 
Calls Received
    & \textit{No calls received} 
        & \textit{$\geq 1$ calls received} \\ 
\rowcolor[HTML]{E0E0E0} 
SMS Messages Sent
    & \textit{No SMS messages sent} 
        & \textit{$\geq 1$ messages sent} \\ 
SMS Messages Received
    & \textit{No SMS messages received}
        & \textit{$\geq 1$ messages received} \\ 
\rowcolor[HTML]{E0E0E0} 
Conversations Detected
    & \textit{No detected conversations} 
        & \textit{$\geq 1$ detected conversations} \\ 
\end{tabular}
}
\end{table}
\vspace{-7mm}

\subsection{Network Structure Estimation \& Description}
Although participant data is initially chronologically ordered, filtering according to a selected behavioral category, either social isolation or sociability, results in time-series segmentation. This segmentation necessitates the identification of a statistical model that can account for this lack of temporal consistency. Correlation network models applied to cross-sectional data deal well with this lack of consistency, as they are effective at visualizing relationships between variables at a specific point in time \cite{borsboom2021network}. As a result we can use this model to generate networks for each person that represent an aggregated average for a sample taken from each selected behavioral category. This allows us to compare network structures of EMAs when participants are, for instance, spending time socially isolated (e.g. remaining at home) or engaging socially (e.g. visiting locations outside of their home). Structurally, EMAs are represented as network nodes with edges between nodes visualizing the linear relationship between each. The strength and sign of any given relationship is defined by the correlation coefficient. Numerically, correlations between nodes range between $-1$ and $1$; with $-1$ indicating a perfect negative relationship, $1$ a perfect positive relationship, and $0$ no relationship at all. Pearson's R Correlation Coefficient is used to calculate the associations between all 10 EMA nodes. For each behavioral context two networks are generated, one for each category within a given behavior (Table \ref{tab:features}).

\vspace{-1mm}
\subsection{Permutation Testing}
To discern whether variations in each behavioral context and their observed network structures differ in statistically meaningful ways, permutation testing is used. The goal of which is to evaluate the null hypothesis that these variations have no discernible concurrence with a participant's self-reported EMAs.

\vspace{-4mm}
\begin{figure}
    \centering
    \includegraphics[width=\textwidth]{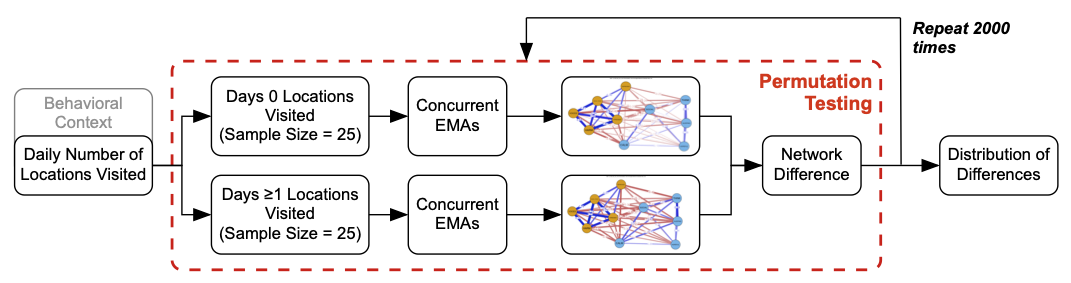}
    \caption{Permutation testing process for \textbf{Daily Number of Locations Visited}, networks are generated using concurrent EMAs from each behavioral category, in turn allowing for the calculation of differences in network connectivity.} \label{fig2:permutation}
\end{figure}
\vspace{-4mm}

Procedurally, permutation testing requires a selected behavior to be filtered according to predefined categories (see Table \ref{tab:features}). A network is generated for each category using a $25$ day sample, at the end of each permutation the observed difference in network connectivity is calculated by subtracting the sum of the network matrices from each other. In probability theory, Central Limit Theorem (CLT) suggests a sample size of approximately $30$, however, to maximise the number of viable participants a reduced sample size of $25$ days is used throughout the permutation testing process. Repeated $2000$ times, a new observed difference in network connectivity is calculated for each permutation, from which a distribution of these differences is then produced. Figure \ref{fig2:permutation} demonstrates this process using the daily number of locations visited as an example.

The resulting distribution of differences in network connectivity for a given behavior can then be compared with a baseline distribution. This baseline undergoes the same testing process but is generated using randomly sampled data, and serves as an empirical distribution from which comparisons can then be made. To measure statistical significance, paired-sample t-testing is used to compare a given behavioral distribution with the empirical baseline distribution. The resulting t-score and p-value can then be used to confirm or reject the null hypothesis that a selected sensor based behavioural context has no discernible influence over an individual's network connectivity.

\vspace{-1mm}
\section{Results}
In this section we present the findings of our analysis; first detailing results for a single participant, and then at a wider level for multiple participants.

\vspace{-1mm}
\subsection{Example of an Individual CrossCheck Participant}\label{sec:ind}
\begin{figure}
     \centering
     \begin{subfigure}[b]{0.49\textwidth}
         \centering
         \includegraphics[width=\textwidth]{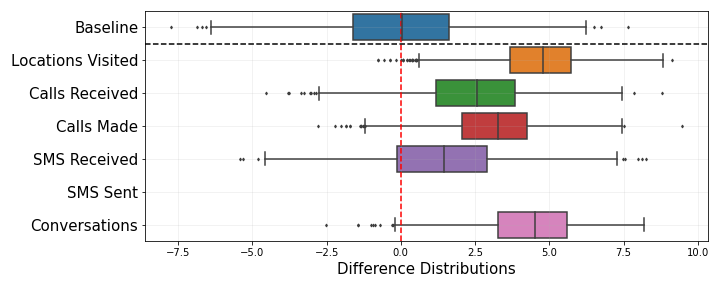}
         \caption{Positive EMAs Only}
         \label{fig:u007:pos:box}
     \end{subfigure}
     \hfill
     \begin{subfigure}[b]{0.49\textwidth}
         \centering
         \includegraphics[width=\textwidth]{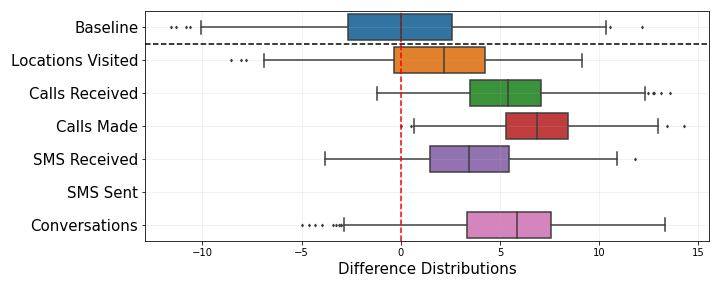}
         \caption{Negative EMAs Only}
         \label{fig:u007:neg:box}
     \end{subfigure}
     \caption{Side-by-side comparison of permutation test results for a single individual across all selected behavioral contexts as well a baseline test - networks generated during testing process used positive EMAs only versus negative EMAs only}
     \label{fig:u007:boxplots}
\end{figure}

Having recorded 330 days of usable data, this individual returned results across 5 of the 6 selected behavioral contexts. Figure \ref{fig:u007:boxplots} visualizes the results produced during permutation testing for each behavior according to networks generated using only positive, and only negative EMAs (see Table \ref{tab:emas}). 

\vspace{-3mm}
\vspace{-5mm}

In each instance, we observe differences when comparing each behavior to a baseline distribution, for example in Figure \ref{fig:u007:pos:box}, we observe a baseline mean of 0 compared to a mean of 4.7 in detected conversations. Focusing in on daily number of locations visited, there is not only an observable difference between behavior and baseline, but there is a notable difference when comparing positive and negative EMAs.

\vspace{-3mm}
\begin{figure}[!ht]
     \centering
     \begin{subfigure}[b]{0.49\textwidth}
         \centering
         \includegraphics[width=\textwidth]{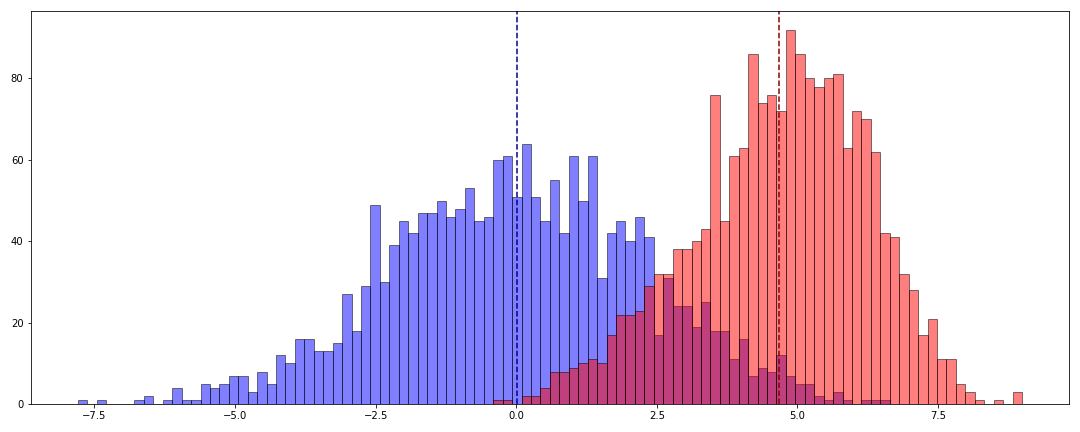}
         \caption{Positive EMAs Only}
         \label{fig:u007:pos:dis}
     \end{subfigure}
     \hfill
     \begin{subfigure}[b]{0.49\textwidth}
         \centering
         \includegraphics[width=\textwidth]{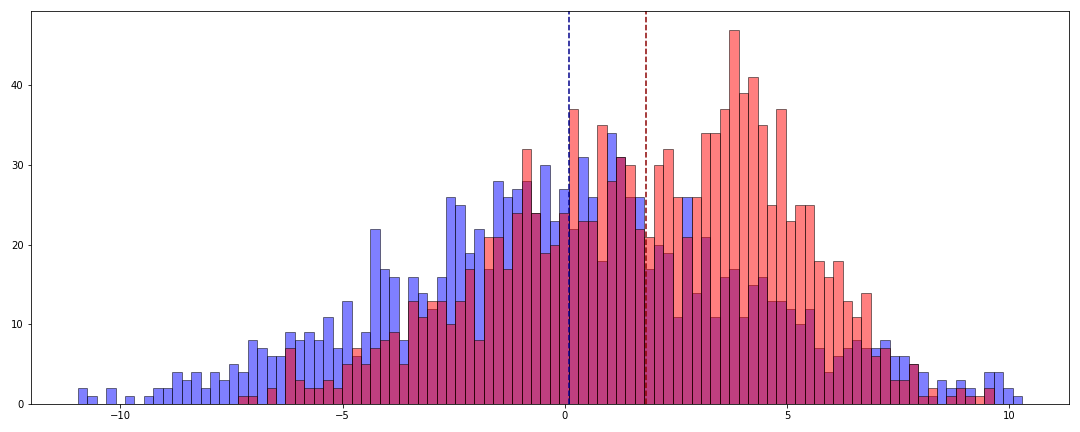}
         \caption{Negative EMAs Only}
         \label{fig:u007:neg:dis}
     \end{subfigure}
     \caption{More detailed analysis of participants results for both their baseline distribution (Blue) and \textbf{Number of Locations Visited} distribution (Red). Fig \ref{fig:u007:pos:dis} clearly shows an observable difference between the baseline distribution and the behavioral context distribution for positive EMAs.}
     \label{fig:u007:dists}
\end{figure}
\vspace{-5mm}

Figure \ref{fig:u007:dists} provides a more detailed visualization of this participant's baseline distribution (in blue) and variations in their daily number of locations visited (in red). Upon visual inspection, there is a clear observable difference in distributions produced using positive EMAs (fig \ref{fig:u007:pos:dis}) when compared to those produced using negative EMAs (fig \ref{fig:u007:neg:dis}). This suggests that, for this participant, there is a noticeable impact on their self-reported positive EMAs when factoring in variations in the daily number of locations visited. Table \ref{tab:p1:results} provides a breakdown of these results across all EMA groups, with a consistent p-value $< 0.001$ indicating these results are statistical significant. Moreover, we see a more sizeable t-score for this participant's positive EMAs ($-72.78$), further suggesting that variations in this behavioral context impacts this individual's self-reporting habits - particularly for their positive EMAs.

\vspace{-3mm}
\begin{table}[!ht]
    \centering
    \caption{Participant's baseline and \textbf{Daily Number of Locations Visited} test results, including mean ($\bar{x}$), standard deviation ($\sigma$), t-score ($t$) and p-value ($p$)}
    \label{tab:p1:results}
    \vspace{3mm}
    \begin{minipage}{0.95\textwidth}
        \resizebox{\textwidth}{!}{%
        \begin{tabular}{L{3.5cm} | C{1.5cm} C{1.5cm} | C{1.5cm} C{1.5cm} C{2.5cm} }
         & \multicolumn{2}{c|}{\textbf{Baseline}} & \multicolumn{3}{c}{\textbf{Daily Number of Locations Visited}}  \\
         & \textbf{$\bar{x}$} & \textbf{$\sigma$} & \textbf{$\bar{x}$} & \textbf{$\sigma$} & \textbf{$t$} \\ \hline
        \rowcolor[HTML]{D9D9D9} 
        All EMAs & -0.44 & 9.06 & 17.30 & 6.22 & -55.62 * \\
        Positive EMAs & 0.05 & 2.28 & 4.61 & 1.55 & -72.72 * \\
        \rowcolor[HTML]{D9D9D9} 
        Negative EMAs & -0.05 & 3.80 & 1.83 & 3.07 & -13.21 *
        \end{tabular}%
        \footnotetext{* $p < 0.001$}
        }%
    \end{minipage}
\end{table}
\vspace{-5mm}

\begin{figure}[]
     \centering
     \begin{subfigure}[b]{0.49\textwidth}
         \centering
         \includegraphics[width=\textwidth]{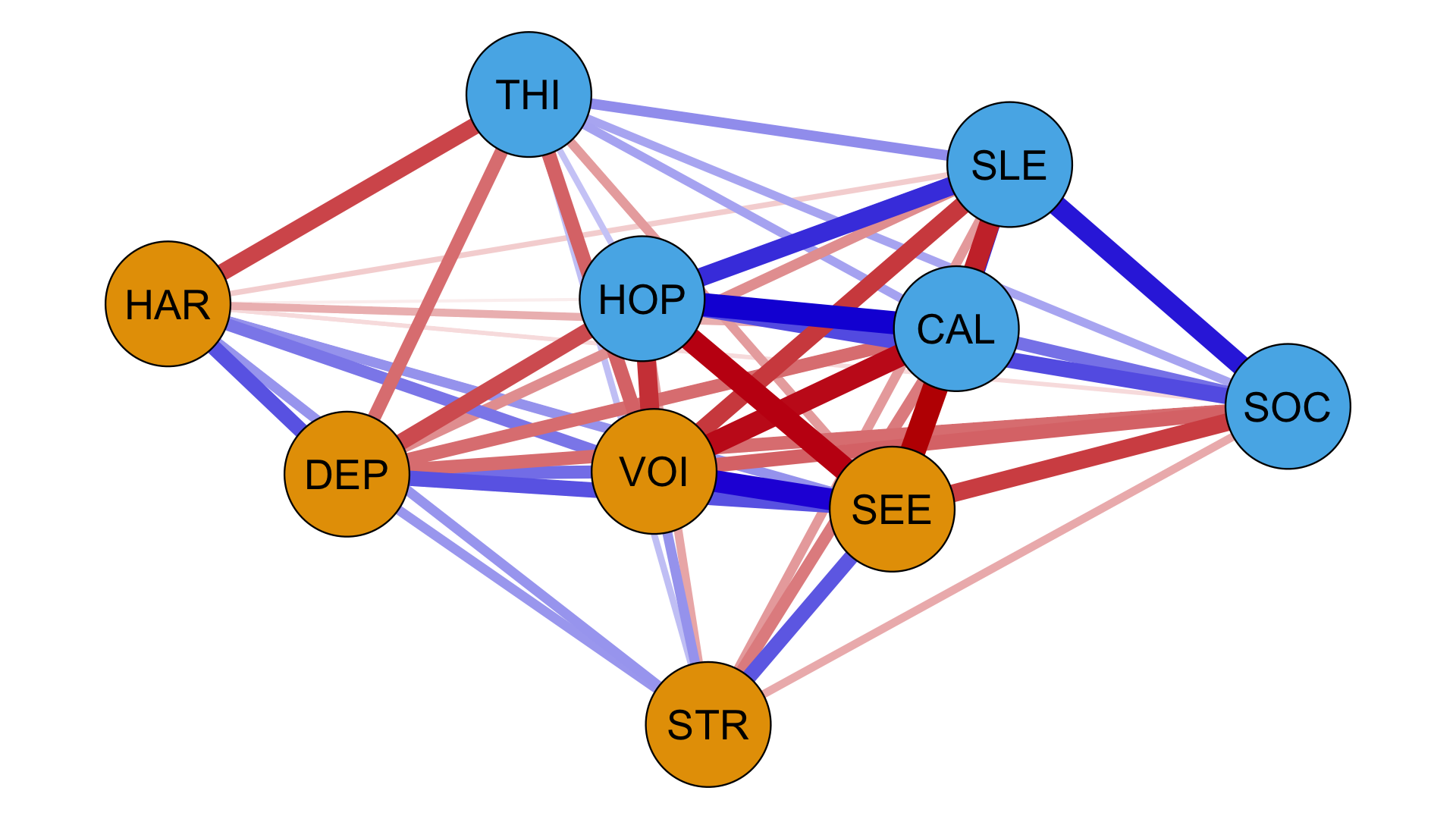}
         \caption{Days with 0 locations visited}
         \label{fig:u007:l=0}
     \end{subfigure}
     \hfill
     \begin{subfigure}[b]{0.49\textwidth}
         \centering
         \includegraphics[width=\textwidth]{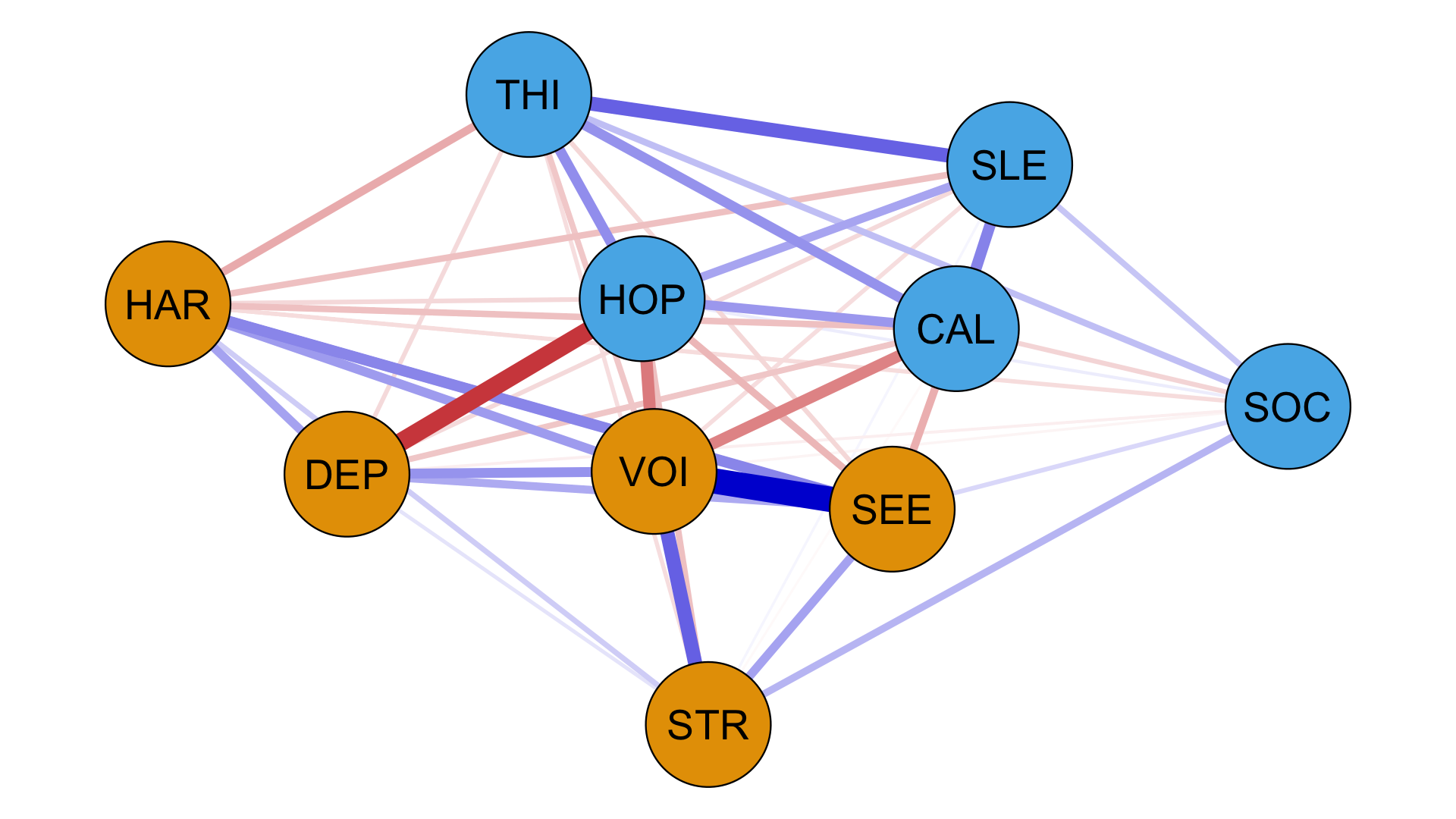}
         \caption{Days with 1 or more locations visited}
         \label{fig:u007:l>0}
     \end{subfigure}
     \caption{Cross-Sectional Networks for \textbf{Daily Number of Locations Visited}. Nodes: \textbf{DEP}ressed, \textbf{HAR}m, \textbf{SEE}ing Things, \textbf{STR}essed, Hearing \textbf{VOI}ces, \textbf{CAL}m, \textbf{HOP}e, \textbf{SLE}ep, \textbf{SOC}ial, \textbf{THI}nking Clearly. Thicker edges denote stronger relationships with blue edges indicating positive correlations and red negative. As expected, one can observe positive relations between positive EMAs (i.e., when this person reports one positive EMA, they are more likely to report others, too); positive relations between negative EMAs; and negative relations between positive and negative EMAs.}

     \label{fig:u007:networks}
\end{figure}

Whilst numerically these results indicate a statistical significance for this particular behavioral context; network generation provides a visualization of the relationship between EMAs. Figure \ref{fig:u007:networks} provides a side-by-side comparison between two networks that each visualize one of the two categories for variations in the daily number of locations this individual participant has visited.

\vspace{-3mm}
\vspace{-5mm}

Both networks in Figure \ref{fig:u007:networks} present strong positive relationships between auditory and visual hallucinations, however, on days where this participant remained at home (see Figure \ref{fig:u007:l=0}) we see a much more complex network with the presence of stronger edges in greater numbers. A visual analysis of these two networks suggests that, for this particular participant, there is a beneficial link between social engagement (interacting with locations away from home) and improvements in self-reported EMAs. In particular, Figure \ref{fig:u007:l=0} illustrates strong negative relationships between each hallucinatory node and feelings of calm and hopeful, this suggests that increased instances of one has a detrimental impact on the other. For example, hallucinations experienced at home could have a more noticeable detrimental impact on this participant's ability to feel calm and hopeful; likewise it could also suggest that feeling calm within this participant's own home reduces the likelihood of them experiencing increased hallucinations.

\vspace{-1mm}
\subsection{By Behavioral Context}
In the interest of brevity the following results focus on the analysis of all 10 EMAs for three specific behavioral contexts. The daily number of calls made, daily number of conversations detected and daily number of locations visited.

\begin{figure}[!ht]
     \centering
     \begin{subfigure}[b]{0.32\textwidth}
         \centering
         \includegraphics[width=\textwidth]{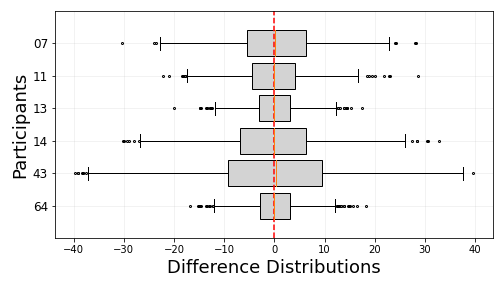}
         \caption{Baseline}
         \label{fig:multi:con}
     \end{subfigure}
     \hfill
     \begin{subfigure}[b]{0.32\textwidth}
         \centering
         \includegraphics[width=\textwidth]{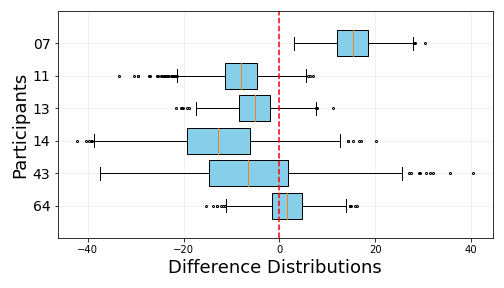}
         \caption{Calls Made}
         \label{fig:multi:call_out}
     \end{subfigure}
     \hfill
     \begin{subfigure}[b]{0.32\textwidth}
         \centering
         \includegraphics[width=\textwidth]{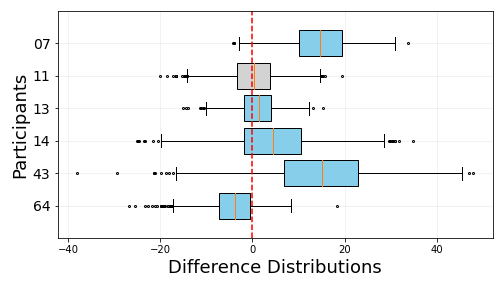}
         \caption{Conversations}
         \label{fig:multi:convo}
     \end{subfigure}
     \hfill
     \caption{Results across for \textbf{all 10 EMAs} for participants who returned results for both selected behavioral contexts - \textbf{Daily Number of Calls Made} and \textbf{Daily Number of Detected Conversations}. Plot color varies depending on statistical significance when using pair-sampled t-testing to compare with the baseline distribution - blue if $p < 0.05$ otherwise plot remains grey.}
     \label{fig:multi:6}
\end{figure}

\paragraph{Daily Number of Calls Made \& Detected Conversations:} Figure \ref{fig:multi:6} presents the distributions of 6 participants who each returned results for both the daily number of calls made and number of conversations detected.

As expected, the baseline distributions in fig \ref{fig:multi:con} indicate a mean close to $0$, demonstrating the consistency of randomized sampling. However following the same process as with the previous individual's results, a comparison between each participant's baseline distribution and behavioral context distribution demonstrates varying degrees of difference - further highlighting the individuality of participant data.

\begin{table}[!ht]
\centering
\caption{Participant results for paired-sample t-testing between their baseline distribution and \textbf{both} Number of Calls Received and Calls Made. For each behavior results include; mean ($\bar{x}$), standard deviation ($\sigma$), t-score ($t$), and p-value ($p$).}\label{tab:all_results}
\vspace{2mm}
\begin{minipage}{\textwidth}
\centering
\resizebox{\textwidth}{!}{%
\begin{tabular}{ C{1.5cm} | C{1.5cm} C{1.5cm} | C{1.5cm} C{1.5cm} C{2cm} | C{1.5cm} C{1.5cm} C{2cm}}

     & \multicolumn{2}{c|}{\textbf{Baseline}} 
        & \multicolumn{3}{c|}{\textbf{Daily Number of Calls Made}} 
            & \multicolumn{3}{c}{\textbf{Daily Number of Conversations}} \\
\textbf{ID} 
    & \textbf{$\bar{x}$}
        & \textbf{$\sigma$}
            & \textbf{$\bar{x}$}
                & \textbf{$\sigma$} 
                    & $t$
                        & \textbf{$\bar{x}$}
                            & \textbf{$\sigma$} 
                                & $t$ \\ 
\hline
\rowcolor[HTML]{D9D9D9} 
07 & -0.18 & 8.75 & 15.45 & 4.97 & -49.65 *& 14.67 & 6.57 & -46.80 * \\
11 & -0.02 & 6.81 & -8.47 & 5.43 & 44.06 * & 0.24 & 5.41 & -1.24 \\
\rowcolor[HTML]{D9D9D9} 
13 & -0.01 & 5.33 & -4.73 & 4.90 & 19.93 * & 1.00 & 4.09 & -4.91 * \\
14 & -0.27 & 10.62 & -12.40 & 9.58 & 39.43 * & 3.88 & 9.51 & -13.01 * \\
\rowcolor[HTML]{D9D9D9} 
43 & -0.42 & 13.63 & -5.45 & 12.12 & 12.91 * & 14.43 & 11.44 & -37.31 * \\
64 & -0.11 & 4.88 & 1.55 & 4.61 & -9.79 * & -4.16 & 5.00 & 23.47 *   
\end{tabular}%
}%
\footnotetext{* $p < 0.001$} %
\end{minipage}
\end{table}

Table \ref{tab:all_results} provides a breakdown of results for each participant visualized in Figure \ref{fig:multi:6}. Whilst in most cases analysis of these two behaviors produces a statistically significant p-value ($p < 0.001$), there is one instance where this is not the case - participant 11 for daily number of detected conversations. A population level analysis, or an unfiltered analysis would have obscured this outlier individual given the variability in distributions across each participant. Moreover participant 07, the same participant analysed previously (see section \ref{sec:ind}), presents significant differences for each of these behavioral contexts; further strengthening the hypothesis that for this participant activities associated with sociability and social engagement have a marked influence on their self-reported EMAs.

\paragraph{Daily Number of Locations Visited:} The following results are from 8 participant's who returned distributions for this behavioral context across all 10 EMAs. 

\vspace{-3mm}
\begin{figure}[!ht]
     \centering
     \begin{subfigure}[b]{0.49\textwidth}
         \centering
         \includegraphics[width=\textwidth]{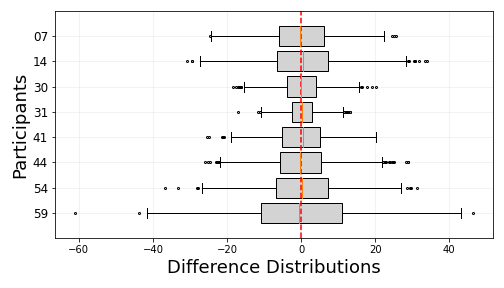}
         \caption{Baseline}
         \label{fig:all:loc_base}
     \end{subfigure}
     \hfill
     \begin{subfigure}[b]{0.49\textwidth}
         \centering
         \includegraphics[width=\textwidth]{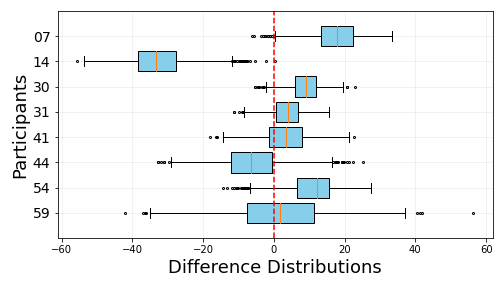}
         \caption{Locations Visited}
         \label{fig:all:loc}
     \end{subfigure}
     \caption{Results across for \textbf{all 10 EMAs} for participants who only returned results for \textbf{Daily Number of Locations Visited}. Plot color varies depending on statistical significance when using pair-sampled t-testing to compare with the baseline distribution - blue if $p < 0.05$ otherwise plot remains grey.}
     \label{fig:all:multiloc}
\end{figure}
\vspace{-5mm}

The side-by-side plots in Figure \ref{fig:all:loc_base} visualize both the baseline and behavioral context distributions for each valid participant, again, we observe statistical significance ($p < 0.05$) across each individual. Table \ref{tab:multi_locations} provides a numerical breakdown of these distributions.

\vspace{-4mm}
\begin{table}[!ht]
\centering
\caption{Results across for \textbf{all 10 EMAs} for participants who returned results for \textbf{Daily Number of Locations Visited}. Results include; mean ($\bar{x}$), standard deviation ($\sigma$), t-score ($t$), and p-value ($p$).}\label{tab:multi_locations}
\vspace{2mm}
\begin{minipage}{\textwidth}
\centering
\resizebox{0.9\textwidth}{!}{%
\begin{tabular}{C{1.5cm} | C{2cm} C{2cm} | C{2cm} C{2cm} C{2.5cm} }
    
    & \multicolumn{2}{c|}{\textbf{Baseline}} 
        & \multicolumn{3}{c}{\textbf{Daily Number of Locations Visited}} \\
\textbf{ID} 
    & $\bar{x}$
        & $\sigma$
            & $\bar{x}$
                & $\sigma$
                    & $t$ \\ 
\hline
\rowcolor[HTML]{E0E0E0} 
07 & -0.07 & 8.76 & 17.42 & 6.40 & -55.33 * \\
14 & 0.44 & 10.28 & -32.71 & 8.16 & 112.87 * \\
\rowcolor[HTML]{E0E0E0} 
30 & 0.09 & 5.60 & 8.96 & 4.15 & -56.95 * \\
31 & 0.10 & 4.43 & 3.80 & 4.39 & -21.34 * \\
\rowcolor[HTML]{E0E0E0} 
41 & -0.03 & 7.84 & 3.34 & 7.14 & -6.05 * \\
44 & -0.09 & 8.21 & -6.26 & 8.56 & 22.57 * \\
\rowcolor[HTML]{E0E0E0} 
54 & 0.09 & 10.32 & 10.81 & 6.80 & -35.19 * \\
59 & 0.13 & 16.00 & 1.54 & 13.96 & -2.25 **
\end{tabular}%
}
\footnotetext{* $p < 0.001$, ** $p < 0.05$} %
\end{minipage}
\end{table}
\vspace{-4mm}

As with previous behavioral contexts, the results presented in Table \ref{tab:multi_locations} further demonstrate the unique behavioral patterns of each participant. Whilst in certain participants we observe a reduced difference between distributions, participant 14 returns a significantly larger result ($t = 112.87$), suggesting that visiting locations outside of the home has an influence over how this individual reports their EMAs. Continued analysis of each behavioral context at a positive and negative EMA level would yield insights into whether or not this influence is specific to one set of EMAs more so then the other.

\vspace{-1mm}
\section{Discussion}
This paper presents an application of an n-of-1 network analysis, leveraging qualitative EMA data whilst factoring in changes and differences in behavioral context measured via sensor data. Specifically, our method allows researchers and clinicians to study, in both exploratory and confirmatory ways, to which degree mental health variables collected via EMA as well as the relation among variables in networks differ across situations. As such, the proposed method provides an inroad to combining multimodal data sources in clinical research and practice, with the goal to enable the potential development of bespoke treatment/management pathways. As an example, the results produced in Figure \ref{fig:u007:dists} reveal that, for this participant at least, not leaving the home significantly changes the way this person reports on his/her positive EMA questions. This new insight coupled with the structure of generated networks could give qualitative actionable information, enabling timely and adaptive interventions for this person's care going forward \cite{nahum2018just}. The next step from this approach will be to expand this methodology and to represent this information in a format this is comprehensible and explainable to the larger psychiatry community.

In summary, the task of how to meaningfully analyse multimodal behavioral sensor data with a complex array of concurrent qualitative self reported data is not well understood, particularly for SMI applications. In this analysis we demonstrate an n-of-1 network analysis approach applied solely to self-reported contextual behavioral data. Networks generated from these distinct periods of behavioral context reveal differences in self-reporting habits, differences beyond chance. This is a first stage indicative approach for datasets similar to CrossCheck which is computationally inexpensive, easily deployed and may lead to actionable clinical insights. However further studies are required to better understand how such insights can be utilized in practice, which are clinically effective but that are also compliant of ethical, regulator and legal requirements.

\vspace{-1mm}
%
%
%
\bibliographystyle{splncs04}
\bibliography{bib.bib}

\begin{thebibliography}{10}
\providecommand{\url}[1]{\texttt{#1}}
\providecommand{\urlprefix}{URL }
\providecommand{\doi}[1]{https://doi.org/#1}

\bibitem{american1994diagnostic}
American Psychiatric~Association, A., Association, A.P., et~al.: Diagnostic and
  statistical manual of mental disorders: DSM-IV, vol.~4. American psychiatric
  association Washington, DC (1994)

\bibitem{american2013diagnostic}
American Psychiatric~Association, D., Association, A.P., et~al.: Diagnostic and
  statistical manual of mental disorders: DSM-5, vol.~5. American psychiatric
  association Washington, DC (2013)

\bibitem{ascher2010cost}
Ascher-Svanum, H., Zhu, B., Faries, D.E., Salkever, D., Slade, E.P., Peng, X.,
  Conley, R.R.: The cost of relapse and the predictors of relapse in the
  treatment of schizophrenia. BMC psychiatry  \textbf{10}, ~1--7 (2010)

\bibitem{bak2016n}
Bak, M., Drukker, M., Hasmi, L., van Os, J.: An n= 1 clinical network analysis
  of symptoms and treatment in psychosis. PloS one  \textbf{11}(9),  e0162811
  (2016)

\bibitem{barnett2018relapse}
Barnett, I., Torous, J., Staples, P., Sandoval, L., Keshavan, M., Onnela, J.P.:
  Relapse prediction in schizophrenia through digital phenotyping: a pilot
  study. Neuropsychopharmacology  \textbf{43}(8),  1660--1666 (2018)

\bibitem{beard2016network}
Beard, C., Millner, A.J., Forgeard, M.J., Fried, E.I., Hsu, K.J., Treadway,
  M.T., Leonard, C.V., Kertz, S., Bj{\"o}rgvinsson, T.: Network analysis of
  depression and anxiety symptom relationships in a psychiatric sample.
  Psychological medicine  \textbf{46}(16),  3359--3369 (2016)

\bibitem{ben2017crosscheck}
Ben-Zeev, D., Brian, R., Wang, R., Wang, W., Campbell, A.T., Aung, M.S.,
  Merrill, M., Tseng, V.W., Choudhury, T., Hauser, M., et~al.: Crosscheck:
  Integrating self-report, behavioral sensing, and smartphone use to identify
  digital indicators of psychotic relapse. Psychiatric rehabilitation journal
  \textbf{40}(3), ~266 (2017)

\bibitem{benoit2020systematic}
Benoit, J., Onyeaka, H., Keshavan, M., Torous, J.: Systematic review of digital
  phenotyping and machine learning in psychosis spectrum illnesses. Harvard
  Review of Psychiatry  \textbf{28}(5),  296--304 (2020)

\bibitem{birchwood2000schizophrenia}
Birchwood, M., Spencer, E., McGovern, D.: Schizophrenia: early warning signs.
  Advances in Psychiatric Treatment  \textbf{6}(2),  93--101 (2000)

\bibitem{borsboom2021network}
Borsboom, D., Deserno, M.K., Rhemtulla, M., Epskamp, S., Fried, E.I., McNally,
  R.J., Robinaugh, D.J., Perugini, M., Dalege, J., Costantini, G., et~al.:
  Network analysis of multivariate data in psychological science. Nature
  Reviews Methods Primers  \textbf{1}(1), ~58 (2021)

\bibitem{bradbury2020practice}
Bradbury, J., Avila, C., Grace, S.: Practice-based research in complementary
  medicine: could n-of-1 trials become the new gold standard? In: Healthcare.
  vol.~8, p.~15. MDPI (2020)

\bibitem{brown2022digital}
Brown, L.A., Taylor, D.J., Bryan, C., Wiley, J.F., Pruiksma, K., Khazem, L.,
  Baker, J.C., Young, J., O'Leary, K.: Digital phenotyping to improve
  prediction of suicidal urges in treatment: study protocol. Aggression and
  violent behavior  \textbf{66},  101733 (2022)

\bibitem{canas2023counterfactual}
Canas, J.S., Gomez, F., Costilla-Reyes, O.: Counterfactual explanations and
  predictive models to enhance clinical decision-making in schizophrenia using
  digital phenotyping. arXiv preprint arXiv:2306.03980  (2023)

\bibitem{chalmers1981method}
Chalmers, T.C., Smith~Jr, H., Blackburn, B., Silverman, B., Schroeder, B.,
  Reitman, D., Ambroz, A.: A method for assessing the quality of a randomized
  control trial. Controlled clinical trials  \textbf{2}(1),  31--49 (1981)

\bibitem{chong2016global}
Chong, H.Y., Teoh, S.L., Wu, D.B.C., Kotirum, S., Chiou, C.F., Chaiyakunapruk,
  N.: Global economic burden of schizophrenia: a systematic review.
  Neuropsychiatric disease and treatment pp. 357--373 (2016)

\bibitem{davidson2022crossroads}
Davidson, B.I.: The crossroads of digital phenotyping. General Hospital
  Psychiatry  \textbf{74},  126--132 (2022)

\bibitem{emsley2013nature}
Emsley, R., Chiliza, B., Asmal, L., Harvey, B.H.: The nature of relapse in
  schizophrenia. BMC psychiatry  \textbf{13}, ~1--8 (2013)

\bibitem{fisher2018lack}
Fisher, A.J., Medaglia, J.D., Jeronimus, B.F.: Lack of group-to-individual
  generalizability is a threat to human subjects research. Proceedings of the
  National Academy of Sciences  \textbf{115}(27),  E6106--E6115 (2018)

\bibitem{fonseca2022risk}
Fonseca-Pedrero, E., Al-Halab{\'\i}, S., P{\'e}rez-Alb{\'e}niz, A.,
  Debban{\'e}, M.: Risk and protective factors in adolescent suicidal
  behaviour: a network analysis. International journal of environmental
  research and public health  \textbf{19}(3), ~1784 (2022)

\bibitem{he2020assessing}
He-Yueya, J., Buck, B., Campbell, A., Choudhury, T., Kane, J.M., Ben-Zeev, D.,
  Althoff, T.: Assessing the relationship between routine and schizophrenia
  symptoms with passively sensed measures of behavioral stability. NPJ
  schizophrenia  \textbf{6}(1), ~35 (2020)

\bibitem{hevey2018network}
Hevey, D.: Network analysis: a brief overview and tutorial. Health Psychology
  and Behavioral Medicine  \textbf{6}(1),  301--328 (2018)

\bibitem{insel2017digital}
Insel, T.R.: Digital phenotyping: technology for a new science of behavior.
  Jama  \textbf{318}(13),  1215--1216 (2017)

\bibitem{insel2018digital}
Insel, T.R.: Digital phenotyping: a global tool for psychiatry. World
  Psychiatry  \textbf{17}(3), ~276 (2018)

\bibitem{jacobson2022digital}
Jacobson, N.C., Feng, B.: Digital phenotyping of generalized anxiety disorder:
  using artificial intelligence to accurately predict symptom severity using
  wearable sensors in daily life. Translational Psychiatry  \textbf{12}(1),
  ~336 (2022)

\bibitem{jacobson2020digital}
Jacobson, N.C., Summers, B., Wilhelm, S.: Digital biomarkers of social anxiety
  severity: digital phenotyping using passive smartphone sensors. Journal of
  medical Internet research  \textbf{22}(5),  e16875 (2020)

\bibitem{kamath2022digital}
Kamath, J., Barriera, R.L., Jain, N., Keisari, E., Wang, B.: Digital
  phenotyping in depression diagnostics: Integrating psychiatric and
  engineering perspectives. World Journal of Psychiatry  \textbf{12}(3), ~393
  (2022)

\bibitem{lillie2011n}
Lillie, E.O., Patay, B., Diamant, J., Issell, B., Topol, E.J., Schork, N.J.:
  The n-of-1 clinical trial: the ultimate strategy for individualizing
  medicine? Personalized medicine  \textbf{8}(2),  161--173 (2011)

\bibitem{van2008visualizing}
Van~der Maaten, L., Hinton, G.: Visualizing data using t-sne. Journal of
  machine learning research  \textbf{9}(11) (2008)

\bibitem{mccutcheon2020schizophrenia}
McCutcheon, R.A., Marques, T.R., Howes, O.D.: Schizophrenia—an overview. JAMA
  psychiatry  \textbf{77}(2),  201--210 (2020)

\bibitem{melcher2020digital}
Melcher, J., Hays, R., Torous, J.: Digital phenotyping for mental health of
  college students: a clinical review. BMJ Ment Health  \textbf{23}(4),
  161--166 (2020)

\bibitem{mohr2020digital}
Mohr, D.C., Shilton, K., Hotopf, M.: Digital phenotyping, behavioral sensing,
  or personal sensing: names and transparency in the digital age. NPJ digital
  medicine  \textbf{3}(1), ~45 (2020)

\bibitem{mohr2017personal}
Mohr, D.C., Zhang, M., Schueller, S.M.: Personal sensing: understanding mental
  health using ubiquitous sensors and machine learning. Annual review of
  clinical psychology  \textbf{13},  23--47 (2017)

\bibitem{morriss2013training}
Morriss, R., Vinjamuri, I., Faizal, M.A., Bolton, C.A., McCarthy, J.P.:
  Training to recognise the early signs of recurrence in schizophrenia.
  Cochrane Database of Systematic Reviews  (2013)

\bibitem{nahum2018just}
Nahum-Shani, I., Smith, S.N., Spring, B.J., Collins, L.M., Witkiewitz, K.,
  Tewari, A., Murphy, S.A.: Just-in-time adaptive interventions (jitais) in
  mobile health: key components and design principles for ongoing health
  behavior support. Annals of Behavioral Medicine  \textbf{52}(6),  446--462
  (2018)

\bibitem{national2014psychosis}
{National Collaborating Centre for Mental Health (UK and others)}: Psychosis
  and schizophrenia in adults: treatment and management. London: National
  Collaborating Centre for Mental Health. doi  (2014)

\bibitem{onnela2021opportunities}
Onnela, J.P.: Opportunities and challenges in the collection and analysis of
  digital phenotyping data. Neuropsychopharmacology  \textbf{46}(1),  45--54
  (2021)

\bibitem{patel2014schizophrenia}
Patel, K.R., Cherian, J., Gohil, K., Atkinson, D.: Schizophrenia: overview and
  treatment options. Pharmacy and Therapeutics  \textbf{39}(9), ~638 (2014)

\bibitem{perez2021wearables}
Perez-Pozuelo, I., Spathis, D., Clifton, E.A., Mascolo, C.: Wearables,
  smartphones, and artificial intelligence for digital phenotyping and health.
  In: Digital Health, pp. 33--54. Elsevier (2021)

\bibitem{punja2016n}
Punja, S., Bukutu, C., Shamseer, L., Sampson, M., Hartling, L., Urichuk, L.,
  Vohra, S.: N-of-1 trials are a tapestry of heterogeneity. Journal of Clinical
  Epidemiology  \textbf{76},  47--56 (2016)

\bibitem{rhemtulla2016network}
Rhemtulla, M., Fried, E.I., Aggen, S.H., Tuerlinckx, F., Kendler, K.S.,
  Borsboom, D.: Network analysis of substance abuse and dependence symptoms.
  Drug and alcohol dependence  \textbf{161},  230--237 (2016)

\bibitem{saha2005systematic}
Saha, S., Chant, D., Welham, J., McGrath, J.: A systematic review of the
  prevalence of schizophrenia. PLoS medicine  \textbf{2}(5), ~e141 (2005)

\bibitem{silver2019smartphone}
Silver, L.: Smartphone ownership is growing rapidly around the world, but not
  always equally  (2019)

\bibitem{torous2016new}
Torous, J., Kiang, M.V., Lorme, J., Onnela, J.P., et~al.: New tools for new
  research in psychiatry: a scalable and customizable platform to empower data
  driven smartphone research. JMIR mental health  \textbf{3}(2),  e5165 (2016)

\bibitem{wander2020schizophrenia}
Wander, C.: Schizophrenia: opportunities to improve outcomes and reduce
  economic burden through managed care. Am J Manag Care  \textbf{26},  S62--S68
  (2020)

\bibitem{wang2016crosscheck}
Wang, R., Aung, M.S., Abdullah, S., Brian, R., Campbell, A.T., Choudhury, T.,
  Hauser, M., Kane, J., Merrill, M., Scherer, E.A., et~al.: Crosscheck: Toward
  passive sensing and detection of mental health changes in people with
  schizophrenia. In: 2016 ACM Int. Joint Conf. Pervasive \& Ubiquitous Comput.
  pp. 886--897 (2016)

\bibitem{wang2017predicting}
Wang, R., Wang, W., Aung, M.S., Ben-Zeev, D., Brian, R., Campbell, A.T.,
  Choudhury, T., Hauser, M., Kane, J., Scherer, E.A., et~al.: Predicting
  symptom trajectories of schizophrenia using mobile sensing. Proceedings of
  the ACM on Interactive, Mobile, Wearable and Ubiquitous Technologies
  \textbf{1}(3),  1--24 (2017)

\bibitem{wang2020social}
Wang, W., Mirjafari, S., Harari, G., Ben-Zeev, D., Brian, R., Choudhury, T.,
  Hauser, M., Kane, J., Masaba, K., Nepal, S., et~al.: Social sensing:
  assessing social functioning of patients living with schizophrenia using
  mobile phone sensing. In: Proceedings of the 2020 CHI conference on human
  factors in computing systems. pp. 1--15 (2020)

\end{thebibliography}

\end{document}